# Computer applications in clinical psychology

**Alina Oana Zamoşteanu**
"Tibiscus" University of Timişoara, România

**ABSTRACT.** The computer-assisted analysis is not currently a novelty, but a necessity in all areas of psychology. A number of studies that examine the limits of the computer assisted and analyzed interpretations, also its advantages. A series of studies aim to assess how the computer assisting programs are able to establish a diagnosis referring to the presence of certain mental disorders. We will present the results of one computer application in clinical psychology regarding the assessment of Theory of Mind capacity by animation.

**Introduction**

The implications of the computer-assisted analysis in psychology are not currently a novelty. The computer-assisted psychological approach has become a necessity in all areas of psychology. During the first period, computer applications were used for processing the data resulted from the assessment of the psychological tests. Later, statistical programs started to be used while nowadays computers are a necessity in terms of processing the results of psychological investigations regardless they refer to establishing the significance level of the results, to making predictions with regard to a behaviour, to the development of new psychological tests in order to meet all the psychometric conditions, etc. Finally, the computer-assisted psychological assessment is widespread in all areas of psychology, be it organizational psychology, legal psychology, educational psychology, clinical psychology and even psychotherapy.

In 1956, Mehl published a monograph that discusses the merits of the statistical decision-making methods (objective) versus clinical methods (subjective). His analysis concludes that the decisions based on objective derivatives are more valid than the judgments based on clinical examination, so with a greater degree of subjectivity. Starting from the assessment of 136 studies, an analysis made by Grove, Zald, Lebow, Snitz, Nelson (2000, apud Butcher, Perry, Atlis) concluded on the superiority of the statistical methods of prediction versus to the clinical methods, situated at 10%.

There are also a number of studies that examine the limits of the computer assisted and analyzed interpretations. Among the mentioned problems there are:
- the difficulty of investigating the internal consistency of interpretation reports;
- insufficient knowledge of the used interpretation system;
- lack of expertise in the area of interest (Moreland, 1985);
- interpretative difficulty in identifying the uniqueness of the tested person;
- tendency to consider the computer assistance as an aim in itself, not merely a means to achieve a final result (Snyder, Widiger, Hoover, 1990).

Lately, the areas in which the computer applications can be used have diversified: in the applied psychology, the assistance programs are becoming more and more diverse. There is currently a tendency to replace pencil-paper type tests with those involving the use of computers. The advantages of this method are given by:
- the maintenance of standard conditions in running the examination;
- the avoidance of any errors that may occur in subjects-type pencil-paper examination;
- the errors avoidance in scoring the items;
- the time necessary for the evaluation is reduced.

Disadvantages that may occur are given by:
- the difficulty encountered by some examined subjects to adapt themselves to the computer-assisted examining situation;
- the tendency to generalize the results and the loss of uniqueness in interpretation;
- the tendency to limit ourselves only to a quantitative interpretation of the data, neglecting the process of qualitative analysis of the gathered information.

The computer assisted technology in the field of psychology aims the following areas of application (Sampson, 1990):
1. Administering psychological tests; the examined subject can input his/her answer through the keyboard or through an optical pen, having the possibility to select the correct variant out of a larger number.
2. Scoring the tests through computer applications;
3. Generating scores for the variables assessed by computer;

4. Narrative interpretation of the report and its generation, available for both the examined subject and the examiner (the narrative report may include a profile of the variables involved in the evaluation);
5. Providing the subject with a CD immediately after the examination, containing the interpretation of the assessment results.

## I. Using computer applications in clinical psychology and psychiatry

The psychiatric scales and some psychological tests widely applicable in the clinical domain benefit from specific programs that allow not only the application, but mostly the correction and rapid interpretation obtained by computer.

There is a series of studies that aim to assess how the computer assisting programs are able to establish a diagnosis referring to the presence of certain mental disorders. Thus, Ross, Swinson, Larkin, Doumani ([R+94]) made a study on the Computerized Diagnostic Interview (C-DIS). The authors evaluated a number of 173 subjects with addictive behaviour (drug abuse), using in the assessment both the C-DIS and the Structured Clinical Interview (DSM III R), applied by clinicians. The concordance between the results obtained using the two methods was analyzed. Except for the antisocial disorders and those caused by substance abuse, the correlation between the results obtained using the two tests was poor (using the Kappa coefficient).

Farrell, Camplair, McCullough ([FCM87]) evaluated the ability of the computerized interview to identify the presence of some disorders in a group of 103 subjects who solicited psychological treatment. In addition to the computerized standardized interview applied to subjects, they were also evaluated by therapists on the basis of an unstructured interview. The results showed a low correlation (r.33) between the difficulties reported by subjects in the evaluation interview conducted via computer and the difficulties identified by therapists through an unstructured interview.

A group of researchers belonging to University of California conducted various studies on the use of virtual reality in assessing mental health problems. Rizzo, a researcher at the "Alzheimer's Disease Research Center" states that "using virtual reality researchers and clinicians can more precisely assess a wider range of behavioural responses than they could with the standard pen and paper psychological test". The aim of the study was to assess spatial cognition or the ability to mentally perform rotations. A total of 60 subjects were tested, men and women, aged between 18 and 34 years. The assessment involved both conventional methods, pencil-paper tests, and the virtual reality. The significant differences between men and women determined by the pencil-paper test were not obtained when using the virtual reality, which led the researchers to the conclusion that virtual reality techniques can help subjects to improve spatial skills (rotational skills).The Mental Rotation Test is an important tool for clinicians in diagnosing traumatic brain injuries and distinguishing between Alzheimer's disease and other dementias.

One of the computer applications in clinical psychology is linked to the Theory of Mind - ToM.

In the latest decades, the growing interest in the study of the social cognition, its evolution and explanation focused on the ability to assign new personal mental states to the others. In order to describe the cognitive structures that develop the capacity of mentalising three models were proposed and developed:
1. "Theory-Theory" Perspective regarding ToM,
2. "Theory-Simulation" Perspective regarding ToM,
3. ToM – a modular cognitive mechanism.

1. The "Theory-Theory" Perspective regarding ToM was elaborated by Perner ([Per91]) and is a non-modular model. This model suggests that the human being reaches different stages of representational capacities, during the ontogenesis, starting with the primary representation of the self as acting agent. The secondary representations made after the 2nd year of life allow the discrimination between real and hypothetical situations. Having natural "meta-representations" (similar to scientific theories) allows a person:
- To "theorize" on other persons' representations;
- To have wrong representations.

Unlike ToM strictly modular model, the "theory-theory" model states that the meta-representation system is not restrictively used by ToM.

2. The "simulation" theory stipulates that ToM is directly related to what we call the capacity to predict the other's person behaviour ([DS95]). Unlike the "theory-theory" model, this model assumes that attributing your own mental states is the basis of the inferences about the others' mental states by replicating or miming the mental life of other individuals.

3. ToM - modular cognitive mechanism ([Les87], [Bar95]). The essence of architectural modularity of mind, according to the model proposed by [Fod83], is a set of restrictions in the information flow. A part of the information inside the module is not accessible outside (the so called processing "interlevels") and some external information is not available inside. In other words, we have a set of restrictions that works on the basis of some filters for selecting the information.

According to the first criterion, the central system has no access to the processing interlevels of the module but only to its outputs. The outputs of such a module can be assumed to be representations. According to the second criterion, modularized processes have no access to any external resource or processing. Starting from the modular model of representing the human mind, proposed by Fodor ([Fod83]), Leslie suggests the existence of a separate cognitive module – ToM – "meta-representational model", meaning there is a specialized cognitive mechanism that is competent to provide interpretations of the agents' behaviours because it builds meta-representations.

This mechanism lies in the capacity of acquisition of the theory of mind and can be independently affected / damaged by other cognitive processing systems.

Most studies on the ToM deficit focused on autism ([BLF85]), on patients with frontal lobe lesions ([RB01]) and schizophrenia (Corcoran, Frith, 1996). Clinical research indicates that patients with schizophrenia suffer from social interaction disorders because of their reduced capacity of communication. Frith explains this by the altering of ToM in schizophrenia which translates into the failure of understanding / controlling the own and the others' mental states, respectively associated behaviours.

In 1992, Frith raised the following question: Does psychotic symptoms of schizophrenia could be explained by an inaccurate cognitive representation of one's own or others' intentions - namely to a poor ToM?

Some schizophrenic patients instead of considering opinions as subjective representations of reality, they consider them the same with reality, hence the difficulty in distinguishing between subjective and objective, while maintaining the false opinions in the form of delusion.

Another hypothesis, regarding the way ToM works in schizophrenia was made by Abu-Akel in 1999. Contrary to the model mentioned above, he suggested that some schizophrenic patients, particularly those with positive symptoms, may have a "hyper ToM". The patients' way of "over attributing" their or others' intentions, as manifested in delusion may be linked to an over inference. In other words, quantitatively speaking, there could be a supra-generation of assumptions or an over-attribution of mental states.

Abu-Akel suggests the following model for the deficit of ToM:
1. The actual deficit of ToM,
2. Normal ToM, combined with the lack of ability to apply this knowledge / these opinions,
3. Hyper ToM.

## 2. The assessment of ToM capacity by animation (computer application). Animation description

In 1940, Heider and Simmel produced an animation silent film in which the characters are two triangles and a circle that moves and a square that does not move. Their study shows that, as naturally, all people (except for the autistic children) imagine for the animation a kind of social plot where a large triangle is seen as an aggressor. The study shows that the movement of the shapes leads to automatic animist perceptions. The movement of the shapes is something very natural and social. But if the animation speed increases or decreases, the illusion of human appearance decreases.

One of the sequences presented by Abell ([AHF00]) is shown below:

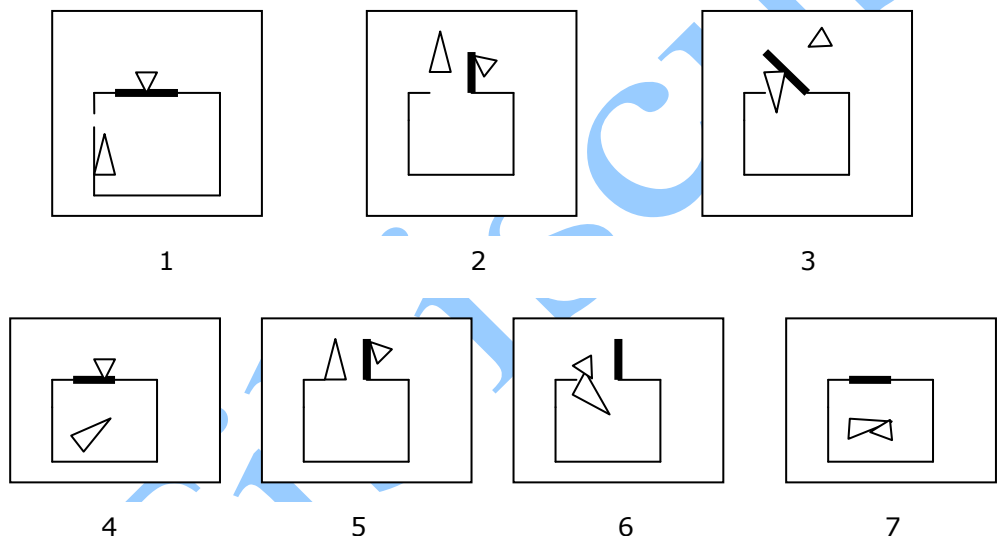

In this animation, the small triangle knocks at the door of the "house" of the big triangle, and then it runs away and hides (pictures 1-5). The big triangle goes to see who is at the door and does not realize that the small one plays a little trick. Between figures 5 and 6 the small triangle makes a "surprise" to the large triangle and in figure 7 the two triangles reconcile.

The instructions used in the application of this animation are:
"I will show you some short films of animation. The characters of these films are some triangles. Please imagine that these triangles can be people and describe as detailed as you can what you think it happens in the movie. You can review each sequence whenever you want".

The subjects' verbal descriptions for each ToM animation are assessed and ranked according to 2 dimensions:

- Accurate "A";
- Intentionality "I".
  In this work we used ([AHF00]):
- 2 animated sequences that present a simple action, with purpose
  - Tracking
    In this sequence the big triangle chases the small triangle.
  - Fight
    In this sequence the two triangles are struggling.
- 3 ToM animated sequences, which present complex interactions, where the action of a triangle influences the action of the other triangle
  - Surprise
    The figure stated above illustrates the "surprise" animation, while providing in the same time also the key descriptive elements.

Scoring criteria for verbal descriptions of animations ([C+00]):

Score (0-5) for Intentionality:

0 = action nondeliberate
1 = deliberate action with no other
2 = deliberate action with another
3 = deliberate action in response to other's action
4 = deliberate action in response to other's mental state
5 = deliberate action with goal of affecting other's mental state

Score (0-3) for Accuracy

0 = no answer, "I don't know"
1 = inappropriate answer: reference to the wrong type of interaction between triangles
2 = partially correct answer: reference to correct type of interaction but confused overall description
3 = appropriate, clear answer

The research was conducted on 35 subjects, aged 30-45 years old, with a primary diagnosis of schizophrenia, aiming the relationship between the severity of symptoms (SCL-90) and ToM capacity (Zamoşteanu and Dedu in 2007).

In this context, we assumed that there is a negative correlation between severity of symptoms in schizophrenia (evaluated by the scale of psychoticism of SCL-90) and ToM capacity.

**Table 1. Correlation coefficients psychoticism - ToM capacity**

|  | p | r |
|---|---|---|
| accuracy film1 | 0.863 | -0.033 |
| accuracy film2 | 0.789 | -0.051 |
| accuracy tom1 | 0.101 | -0.305 |
| accuracy tom2 | 0.049 | -0.362 |
| accuracy tom3 | 0.009 | -0.466 |
| intentionality tom1 | 0.001 | -0.577 |
| intentionality tom2 | 0.0001 | -0.608 |
| intentionality tom3 | 0.025 | -0.407 |

The statistic results show that there is a strong negative correlation ($r_1 = -0.57$ at $p_1 = 0.001$, $r_2 = -0.608$ at $p_2 = 0.0001$, $r_3 = -0.407$ at $p_3 = 0.025$) between the severity of symptoms and the score obtained for the complex animations from the point of view of intentionality. Consequently, it may be state that the more the severity of psychotic symptoms increases, the more the ability of ToM decreases, regarding the expression of intentionality on complex tasks.

For to make a measure of the strength of the relation, we have used the correlation coefficient, using the formula:

$$r^2 = 1 - \frac{SE^2}{SD^2}$$

where: SE = deviation scores
SD = variance

This formula allows us to visualize what r is, but it would be a very awkward and tedious way of actually calculating it. A much easier way is given by the following formula:

$$r = \frac{N \sum XY - \sum X \sum Y}{\sqrt{N \sum X^2 - (\sum X)^2} \sqrt{N \sum Y^2 - (\sum Y)^2}}$$

where X, Y = variables.

In terms of accuracy, 2 of 3 animations, show negative correlation ($r_2 = -0.362$ at $p_2 = 0.049$, $r_3 = -0.466$ at $p_3 = 0.009$) between the severity of symptoms and level of accuracy scores on complex tasks.

The first complex animation, does not show the existence of any significant correlation ($r_1 = -0.305$ at $p_1 = 0.101$) between severity of the symptoms and the level of accuracy. The two animations having a simple purpose show no significant negative correlation (r1 = -0.033 at p1 =0.863, r2 = -0.051 at p2 =0.789) between the severity of symptoms and the accuracy level.

This could mean that in fact the capacity of ToM deteriorates along with the worsening of the psychotic symptoms, regarding the ability of expressing the intentionality and of the understanding the intentionality of the actions with a complex purpose.

The modular theory makes the distinction between abstract mental states, representing a ToM ability and the inference about the physical world, which is considered to be a general cognitive ability. In this case, the results show that subjects are not able to infer about abstract mental states, yet are able to infer about actions with a simple purpose (to understand the simple intentions of the third parties).

Another hypothesis was to find the existence of significant differences regarding ToM capacity in terms of expression and understanding complex actions intentionality. We assumed also that there was no significant difference in understanding simple goal tasks' intentionality among patients diagnosed with schizophrenia and subjects without a psychiatric diagnosis.

**Table 2. Descriptive Statistics**

|  | criterion | N | mean | σ |
|---|---|---|---|---|
| accuracy film1 | 1.000 | 30.000 | 2.467 | 0.730 |
|  | 2.000 | 22.000 | 2.682 | 0.568 |
| accuracy film2 | 1.000 | 30.000 | 2.000 | 0.910 |
|  | 2.000 | 22.000 | 2.227 | 0.813 |
| accuracy tom1 | 1.000 | 30.000 | 2.167 | 0.531 |
|  | 2.000 | 22.000 | 2.682 | 0.477 |
| accuracy tom2 | 1.000 | 30.000 | 1.500 | 0.572 |
|  | 2.000 | 22.000 | 2.545 | 0.510 |
| accuracy tom3 | 1.000 | 30.000 | 1.500 | 0.820 |
|  | 2.000 | 22.000 | 2.364 | 0.727 |
| intentionality tom1 | 1.000 | 30.000 | 2.133 | 0.681 |
|  | 2.000 | 22.000 | 4.000 | 1.024 |
| intentionality tom2 | 1.000 | 30.000 | 2.133 | 0.776 |
|  | 2.000 | 22.000 | 3.273 | 1.120 |
| intentionality tom3 | 1.000 | 30.000 | 2.067 | 0.944 |
|  | 2.000 | 22.000 | 3.318 | 1.086 |

Table 3. The differences between averages

|                    | t      | gl     | p     |
|--------------------|--------|--------|-------|
| accuracy film1     | -1.149 | 50.000 | 0.256 |
| accuracy film2     | -0.947 | 47.989 | 0.348 |
| accuracy tom1      | -3.608 | 50.000 | 0.001 |
| accuracy tom2      | -6.810 | 50.000 | 0.000 |
| accuracy tom3      | -3.934 | 50.000 | 0.000 |
| intentionality tom1| -7.896 | 50.000 | 0.000 |
| intentionality tom2| -4.102 | 35.219 | 0.000 |
| intentionality tom3| -4.430 | 50.000 | 0.000 |

The results show that the null hypothesis is not confirmed.

It can be said that there is no statistically significant difference, so the first task (t = -1.149 for p = 0.256) and the second task (t = -0947 for p = 0.348) among patients diagnosed with schizophrenia and subjects without a psychiatric diagnosis in terms of accuracy, so the understanding of the simple goal tasks' intentionality.

The average scores obtained regarding accuracy in simple goal tasks by schizophrenic subjects (m = 2.467, σ = 0.730 respectively m = 2.00, σ = 0.910) do not differ significantly from the scores' average obtained in accuracy by the subjects without a psychiatric diagnosis (m = 2.682, σ = 0.568, respectively m = 2.227, σ = 0.813).

In terms of understanding and expressing intentionality of complex tasks, statistical results show that (for ToM animations t, p, m and σ will be noted based on the number of animation): there is a statistically significant difference between schizophrenia subjects and those with no psychiatric diagnosis regarding the mean scores obtained for accuracy (t1= -0.608 at p1=0.001, t2 = -6.810 at p2 = 0.0001, t3 = -3.934 at p3 = 0.0001).

The accuracy obtained by the schizophrenic subjects (m1 = 2.167, σ1 = 0.531, m2 = 1.500, σ2 = 0.572 and m3 = 1.500, σ3 = 0.820) differs significantly from the scores' mean obtained by the subjects without psychiatric diagnosis (m1 = 2.682, σ1 = 0.477, m2 = 2.54, σ2 = 0.510 and m3 = 2.364, σ3 = 0.727), the difference being in the favour of the last mentioned subjects.

There is a statistically significant difference between schizophrenia subjects and those with no psychiatric diagnosis in terms of scores' mean obtained on intentionality (t1= -7.896 , t2 = -4.102, t3 = -4.430 la p1 = p2 = p3 = 0.0001).

Scores' mean obtained in intentionality by schizophrenic subjects ($m_1 = 2.133$, $\sigma_1 = 0.681$, $m_2 = 2.133$, $\sigma_2 = 0.776$ and $m_3 = 2.067$, $\sigma_3 = 0.944$) differs significantly from the average scores obtained in accuracy by the subjects without psychiatric diagnosis ($m_1 = 4.00$, $\sigma_1 = 1.024$, $m_2 = 3.273$, $\sigma_2 = 1.120$ and $m_3 = 3.318$, $\sigma_3 = 1.086$), the difference being in the favour of the last mentioned group.

In other words, both subjects without psychiatric diagnosis and those diagnosed with schizophrenia have the same understanding of the goal of simple actions. However, this does not totally exclude the possibility that even this part of ToM capacity may be affected. Although not a criterion, schizophrenic subjects who participated to this study were not selected according to the length of hospitalization, but none had been hospitalized more than 6 years.

Under these circumstances, accepting the Leslie's model (ToM = separate cognitive module) and the assumption that this ToM capacity is rather state than trait, it can be assumed that the deterioration occurs gradually. The problem raised in this context is whether this deterioration occurs in opposite direction to ToM capacity acquisition, or not.

Thus, this ToM capacity may deteriorate in the first phase in terms of understanding and expressing complex tasks intentionality, and later even the basic function of TOM, namely the understanding of simple goal tasks, to be affected.

This comes to further support Leslie's model regarding the ToM modularity, in the sense that the mechanism deterioration that forms the basis of theory of mind acquisition capacity may happen independently of other cognitive processing systems.

The analysis of verbal descriptions of patients pointed out that they were unable to describe a complex animation. Even if they did not understand the animator's intention, nor were able to express a complex intention, however, they described simple motions, actions with simple goal:

- For the ToM animation „Seduction", a subject diagnosed with schizophrenia gave the following description: „the blue triangle got into the square; the red triangle walks; they both became an 8; the blue triangle came out and the red triangle walked";
- For the simple goal animation, „Fight", the same subject provided the following description: „they walk left, in the square, like starting a fight, it is an X as an eight; the red one is more courageous; yes... they fight".

In this description the fact can be observed that the subject could not understand the complex intentionality, but tried to describe it using simple actions.

Another subject diagnosed with schizophrenia said for the animation „Surprise": „They are just some triangles and a square; I cannot imagine anything", and for the „Fight" animation the subject stated: „Two trees that have problems due to wind and fight". Basically, the subject failed to assign any type of complex mental state.

Since the listing of descriptions was made utilising only the used verb, in order to avoid experimenter's bias in interpretation, it is difficult to assess if the subject indeed understood the simple action „fight", when receiving instructions for interpretation.

Another aspect that deserves to be postponed is related to the length of descriptions. Subjects diagnosed with schizophrenia had comparable longer descriptions to simple goal animations than to those containing complex intentional goal. This may be due to characteristics of delirious thinking in psychosis, namely over-inference, and may support the hypothesis that the ToM capacity may suffer deterioration in various psychoses. It can be observed in schizophrenic subjects' descriptions of complex animations that, although they use mentalising verbs and describe simple intentional actions, they are not able to make the connection between these actions. With worsening symptoms the ToM deterioration is present also at the level of the module's basic structures, and is manifested through the inability to understand simple actions in terms of intentionality.

**Conclusions**

Although there are some inconsistencies reported between the computer assisted assessment and other assessment methods (classical methods, the pencil-paper type), we can only state that there is not a perfect union between the two types of assessments. The computer assistance and analysis cannot substitute the other existing methods. For example, certain relationships between symptoms, which are essential in defining a certain disorder, are difficult to measure using the computer ([Fir94]). Sawyer in 1966 says that the best predictions are made when the clinician gathers information deciding how to use it, while the computers are programmed to provide statistical data and to compare it with the one in the database. Finally, we conclude that computer applications in clinical psychology represent an instrument to support the clinical reasoning that cannot replace the clinician's skills.